\documentstyle[secnumtab,epsfig]{fbssuppl}

\def\lsim{\mathrel{\raise2pt\hbox to 8pt{\raise -6pt\hbox{$\sim$}\hss{$<$}}}}

\title{Nuclear Forces and Chiral Theories}
\author{J. L. Friar\instnr{1,2}\thanks{{\it E-mail address:} 
friar@sue.lanl.gov}}
\instlist{Theoretical Division, Los Alamos National Laboratory,
MS B283, Los Alamos, NM 87545, USA \and 
Institute for Nuclear Theory, University of Washington, 
Physics-Astronomy Building, Box 351550, Seattle, WA 98195-1550, USA}

\begin{document}

\maketitle
\begin{abstract}
Recent successes in {\it ab initio} calculations of light nuclei (A=2-6) will be
reviewed and correlated with the dynamical consequences of chiral symmetry. The
tractability of nuclear physics evinced by these results is evidence for that
symmetry. The relative importance of three-nucleon forces, four-nucleon forces,
multi-pion exchanges, and relativistic corrections will be discussed in the
context of effective field theories and dimensional power counting. Isospin 
violation in the nuclear force will also be discussed in this context. 

\end{abstract}

\section{Introduction}

The purview of my talk is chiral (symmetry) aspects of nuclear forces.  In order
to treat this expanding topic, I will ask and answer three questions.  The first
question is:  What is chiral symmetry (CS) and where does it come from?  The
second question is:  What influence does chiral symmetry have on nuclear
physics?  Finally, my last question is:  What are effective field theories, and
why should we be interested in them?  The answer to the first question will
illustrate how QCD plays a significant role even in the low-energy regime
appropriate to nuclear ground and low-lying exited states, where quarks and
gluons are not the most appropriate degrees of freedom to perform dynamical
calculations.  The second answer will discuss how the {\bf symmetries} of QCD
persist when one uses pions and nucleons as the relevant degrees of freedom in a
nucleus, and what those symmetries imply in nuclei (power counting).  The last
question will outline an approach to nuclear physics (chiral perturbation theory
or $\chi$PT) that is very recent and does not yet approach the quantitative
sophistication of traditional nuclear physics.  I will outline a recent
calculation of isospin violation in the nuclear force and try to illustrate why
this approach is superior in many ways to the traditional approach.  Finally, I
will conclude and summarize. 

\section{What is Chiral Symmetry and Why Is It Important?}

We are all aware of QCD.  This theory has made a profound impact on nuclear 
physics, where it has nevertheless produced very few concrete results.  Everyone
talks QCD, but in the low-energy regime for more than one nucleon (traditional
nuclear physics) there are few successes at relating QCD to nuclear structure.
The reasons for this are clear.  The success of QCD was founded at first on
its theoretical structure.  It is a structurally simple theory when written in
terms of quark and gluon fields.  Indeed, the theory manifests its symmetries in
terms of these variables, particularly in the limit of vanishing quark masses. 

One such symmetry\cite{1} is chiral $SU(2)_R \times SU(2)_L$ symmetry, where
``R'' and ``L'' refer to right-handed and left-handed (helicity) quarks
\footnote{A similar chiral symmetry arises in QED if the electron mass is set to
zero.  This symmetry accounts for differences between Mott and Rutherford
back-scattering of electrons from nuclei.  This difference provides a mechanism
for the separation of charge and transverse multipoles.} that separately
transform in a particular way in the QCD Lagrangian.  In other words, those
types of quarks don't talk to each other. The vacuum of our world doesn't share
this symmetry and spontaneous symmetry breaking is the result, which leads to
massless pions (for massless quarks). If the quarks are given a small mass,
these Goldstone bosons derive a small mass, as well.  The resulting $SU(2)_V
\times SU(2)_A$ symmetry has a conserved vector current and partially conserved
axial current (PCAC), which would be conserved if pions were massless. 

The ``simple'' symmetries of QCD pose a problem for nuclear physics, however. We
don't ordinarily describe nuclei in terms of quarks.  Although there is nothing
improper with viewing a nucleus as a large ``container'' filled with quarks and
gluons, our basic description is in terms of ``physical'' degrees of freedom. 
If one bombards any nucleus with low-energy photons (for example), nucleons are
ejected.  At somewhat higher energies pions are emitted.  These ejecta are
clearly the appropriate degrees of freedom for low-energy ($\lsim$ a few hundred
MeV) nuclear physics, as our gedanken experiment shows. 

Chiral symmetry can be expressed in terms of the physical degrees of freedom,
but the description is more complicated and not particularly obvious.
Unfortunately this also means that determining the constraints of CS for a
nucleus is nontrivial and is buried in details of the nuclear force.  Chiral
models (or theories) exist for the various building blocks required for
constructing a nuclear force.  We will discuss these details later.  We can
summarize this section by stating: 

\begin{itemize}
\item Chiral symmetry arises naturally in QCD at the quark level and persists in
nuclear physics dynamics expressed in terms of nucleons and pions. 
\end{itemize}

\section{What Influence does Chiral Symmetry Have on Nuclear Physics?}

The influence of CS (or more generally, QCD) on nuclear physics can be stated on
several levels.  Most are obvious, but they should be stated nonetheless. 

\begin{itemize}
\item The pion has a very small mass.
\item The pion is a pseudoscalar particle.
\item The pion is an isovector particle.
\item Chiral symmetry forbids extremely large $\pi N$ interactions.
\item The large-mass scale, $\Lambda$, associated with QCD is $\sim$ 1 GeV.
\end{itemize}
The first four items involve the pion. They and their consequences have been 
well known for decades (even before QCD). The last item is newer and its 
consequences for nuclear physics are much more subtle.

The first and most obvious of these properties leads to OPEP being the
longest-range part of the nuclear force and plausibly the most important
component.  The second follows from the Goldstone theorem and the properties of
the axial current; it leads to a spin-dependent pion-nucleon interaction. Since
the strong-interaction Hamiltonian must be an overall scalar, the pion's
negative parity must be balanced by another negative parity, which can only come
from a vector (e.g., the pion momentum).  This vector must then be contracted
with a pseudovector, of which only the nucleon spin suffices.  Thus we arise at
the usual $\mbox{\boldmath $\sigma$} \cdot \mbox{\boldmath $q$}_{\pi}$ form of
the $\pi N$ interaction.  When we form OPEP, the resulting spin dependence
becomes a tensor force, an important distinction from the much-more-tractable
central forces. 

These two features account for the OPEP dominance seen in light nuclei.  Most of
the potential energy derives from the tensor force, while OPEP dominates\cite{2}
the potential energy:  $\langle V_{\pi} \rangle / \langle V \rangle \sim$ 75\%. 
This complete dominance is due in part to cancellations, but the overall
importance of OPEP is a consequence of chiral symmetry. 

The isovector nature of the pion is of the utmost importance in meson-exchange 
currents. This area of nuclear physics provided the first unambiguous evidence
for pion degrees of freedom in nuclei. The motion of {\bf charged} particles 
in any system produces a current. This fact and the long range of OPEP 
guarantee the pion a dominant role. Recent work in this area\cite{3} is based on
$\chi$PT and power counting, but we have no space to pursue this very 
interesting topic. 

Another unambiguous demonstration of pion degrees of freedom in nuclei has an
added cachet:  it comes with error bars.  Beginning approximately fifteen years
ago, the Nijmegen group have implemented a sophisticated and successful program
of Phase Shift Analysis (PSA) of the $NN$ interactions.  Their methodology
includes treating all known long-range components of the electromagnetic
interaction, such as Coulomb, magnetic moment, vacuum polarization, etc., as
well as the tail of the $NN$ interaction beyond 1.4 fm, which includes OPEP. The
inner interaction region is treated in a phenomenological fashion.  This allows
an accurate determination of the $\pi N$ coupling constants\cite{4}. In order to
check for systematic errors in the analysis, they also fitted the masses of the
exchanged pions (both charged and neutral) and found: 
$$
m_{\pi^{\pm}} = 139.4(10) \, {\rm MeV}, \eqno (1a)
$$
$$
m_{\pi^{0}} = 135.6(13) \, {\rm MeV}. \eqno (1b)
$$
The small error bars ($\lsim$ 1\%) further demonstrate the importance of OPEP in
the nuclear force.  They are currently investigating the tail of the rest of the
$NN$ interaction. 

A valuable byproduct of this work is the ability to construct potentials by
directly fitting to the data, rather than to phase shifts, and to utilize the
entire $NN$ data base.  Several potential models, such as the Argonne $V_{18}$
model, have been constructed in this way and fit the $NN$ data base far better
than any previous attempts.  One useful corollary of this work\cite{5} is that a
baseline has been set for the triton binding energy ($\sim$ 7.62 MeV) using
local $NN$ potentials.  Nonlocal potential components are mandated by relativity
and these are currently under intensive investigation. 

\subsection{An Opinionated Symmetry}

The constraints of chiral symmetry are most easily stated in the form of
``opinions'', which are detailed below.  Because this symmetry is not realized
in the usual way (viz., comparing matrix elements of operators that should be
(nearly) equal), our tests of CS in nuclei are somewhat indirect.  We choose to
express the results of such tests as:  chiral symmetry and dimensional power
counting have an {\bf opinion} about 

\begin{itemize}
\item the relative size of various components of the two-nucleon force (2Nf);
\item the relative sizes of three-nucleon forces (3Nf) and 2Nf;
\item the relative sizes of four-nucleon forces (4Nf) and 2Nf, $\cdots$ ;
\item the relative sizes of various components of the nuclear electromagnetic
currents, such as the impulse approximation, pion-exchange currents,
heavy-meson-exchange currents, etc.; 
\item the size of relativistic corrections to various parts of the dynamics.
\end{itemize}

These ``opinions'' are the result of dimensional power counting -- expressing 
the results of dynamical calculations in terms of powers of the ratio of an 
average momentum (or energy) component, $\bar{p}$, to $\Lambda$, the large-mass
QCD scale. Thus, $(\bar{p} / \Lambda)^N$ represents the progression in the 
sizes of various operators in the nuclear medium.  There are rules\cite{6} for 
determining $N$ for a given case, allowing comparisons to be made.  Moreover, 
in light nuclei we expect $\bar{p} \sim m_{\pi}c^2$. This estimate is useful 
but should not be taken too literally. 

Power counting is not merely based on our wishes (however strong they may be),
but rather on an analysis\cite{6} of the structure of the (generic) effective
Lagrangian (based on QCD) underlying nuclear physics.  That analysis suggests
that the various (dimensional) couplings are given by powers of $f_{\pi}$, the
pion decay constant ($\sim$93 MeV), $\Lambda$ (discussed above), and
dimensionless constants ($\sim$ 1).  The latter assumption is what makes the
whole scheme quantitative.  Thus, the sizes of individual nuclear operators
are expected to be expressible as $\sim (\bar{p} / \Lambda)^N$.  One last
ingredient is needed in order to make the scheme useful.  If $N$ were negative,
higher-order terms in perturbation theory (such as loops) could become quite
large.  Chiral symmetry prevents that and mandates\cite{6} $N \geq 0$.  This
non-obvious and nontrivial requirement allows a convergent perturbation theory
for low-energy strong-interaction dynamics. Also not obvious is the fact that
this condition is equivalent to ``pair suppression'', an {\it ad hoc} (but
phenomenologically necessary) procedure that historically was used to eliminate
large and unphysical two-pion-exchange forces. 

\begin{figure}[htb]
  \epsfig{file=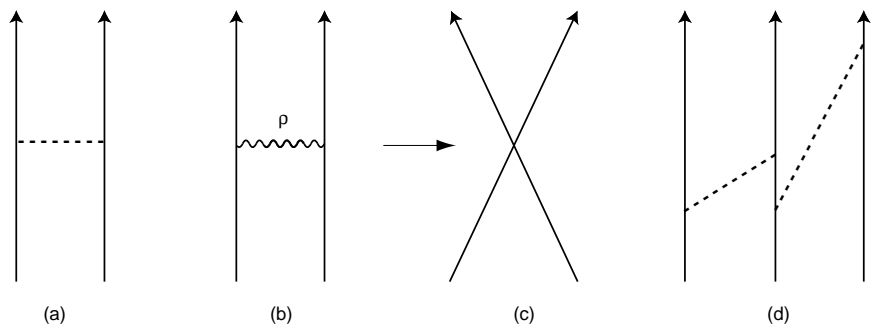,height=2.0in,bbllx=97pt,bblly=435pt,bburx=483pt,bbury=600pt,clip=}
  \caption{Time-ordered perturbation theory diagrams for nuclear potentials
in $\chi$PT, with OPEP shown in (a), $\rho$-exchange in (b) becomes a contact
interaction in (c), while overlapping pion exchanges contribute to the 3Nf in
(d). Pions are depicted by dashed lines, while nucleons are shown as solid 
lines.}
\end{figure}

We can now discuss the various ``opinions'' in turn.  The smallest values of
$N$ for the nuclear force correspond to OPEP (shown in Fig.\ (1a)) and a generic
short-range interaction, illustrated in Fig.\ (1c).  Typically the latter might
arise from $\rho$-exchange, as indicated in Fig.\ (1b), whose range is shrunk to
a point. Massive, unstable particles, resonances, etc., cannot propagate very
far in the low-energy regime appropriate to effective field theories, and their
interaction range is therefore shrunk to a point. Finite-range effects are
introduced as derivatives of zero-range interactions. Higher values of $N$
correspond to $n$-body forces with $N \sim 2n$, and 2-body forces arising from
loops.  A typical (time-ordered) contribution to a three-nucleon force is shown
in Fig.\ (1d).  Thus we expect 4Nf to be smaller than 3Nf, and the latter to be
smaller than 2Nf.  Indeed, the rule of thumb is that adding nucleons irreducibly
to a process increases $N$ by two (each) and adding a loop adds two also. 
Examples of the latter are two-pion-exchange two-nucleon forces and vertex
corrections to one-pion exchange. These forces are therefore considerably weaker
than OPEP. 

We can also ask the obvious question: do models exist that violate the
constraints of CS and generate large 2Nf, 3Nf, $\cdots$ ? The answer
is yes and the problem is indicated in Fig. (2). Pure PS coupling (i.e.,
$\gamma_5$) of a pion to a nucleon leads to large ``pair'' terms as shown in
this figure. These terms are each $\sim (M/m_{\pi})$ larger than chiral models
generate, where $M$ is the nucleon mass. Two such factors are thus $\sim 50$
times larger than what is physically allowed. Very large many-body forces, such
as the 4Nf in Fig. (2b), would severely limit the tractability of nuclear
physics calculations. They are forbidden by the chiral condition, $N \geq 0$.
Thus, CS makes many-nucleon forces small and nuclear physics tractable. 

\begin{figure}[htb]
  \epsfig{file=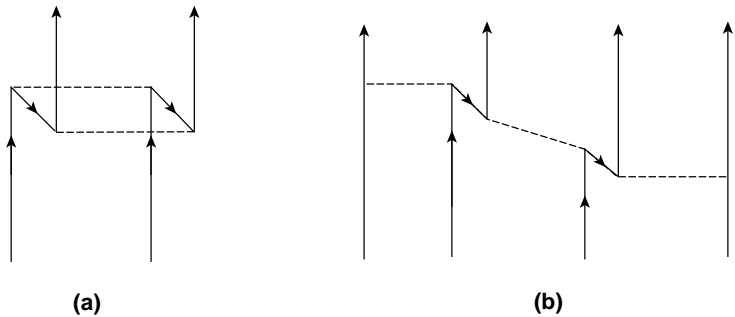,height=2.0in,bbllx=31pt,bblly=436pt,bburx=498pt,bbury=650pt,clip=}
  \caption{Time-ordered perturbation theory diagrams that emphasize ``pair''
contributions to the NN force in (a) and to the 4Nf in (b). Pions are depicted 
by dashed lines, while nucleons are shown as solid lines.}
\end{figure}

Finally, we note that relativistic corrections fit into this scheme, but have
nothing explicitly to do with chiral symmetry.  The question of what determines
$\Lambda$, or even what quantities $\Lambda$ might subsume, has been
deliberately avoided.  We simply note that the nucleon mass, $M$, has a size
$\sim \Lambda$, and that an expansion in powers of $1 / M$ should behave like an
expansion in $1 / \Lambda$ (i.e., it should behave like $(\bar{p} /
\Lambda)^N)$. 

\subsection{Nuclear Few-body Calculations}

Testing these ideas in few-nucleon systems is relatively easy.  Few areas of
nuclear physics have made such substantive progress in the previous 10 years as
few-nucleon physics\cite{7}. We are now solving the Schr\"{o}dinger equation
accurately\cite{8} for the A = 2-6 systems.  Such accurate solutions are usually
called ``exact'' or ``complete'' if their error is $\lsim$1\%.  The emergence of
Green's Function Monte Carlo techniques as the method of choice for nuclear
ground states has made accurate calculations possible for systems that were
thought to be far out of our reach 10 years ago. 

A subset of results from a recent calculation\cite{8} is shown in Table 3.1,
including the approximate date when ``exact'' solutions were first obtained
for each nuclear state.  The calculations were solutions of the Schr\"{o}dinger
equation for a Hamiltonian containing a recent (accurately fit) $NN$ force and a
weak 3Nf, which was adjusted to fit the $^3$H binding energy.  There was no 4Nf.
The agreement between theory and experiment is excellent. 

\begin{table}[hbt]
\centering
\caption{Calculated and experimental ground-state energies (in MeV) of 
    few-nucleon systems, together with (approximate) dates when they were first 
    accurately solved for ``realistic'' potentials.}

\begin{tabular}{|l||llllll|}
\hline
{$^{\rm A}{\rm X} (J^{\pi})$}&\hspace{0.1in} {$^2{\rm H}(1^+)$} & 
{$^3{\rm H}(\frac{1}{2}^+)$} & 
{$^4{\rm He}(0^+)$} &  {$^5{\rm He}(\frac{3}{2}^-)$} & 
{$^5{\rm He}(\frac{1}{2}^-)$} & {$^6{\rm Li}(1^+)$} \\ \hline \hline
{Solved} &\hspace{0.1in}$\sim$1950 & 1984 & 1987 & 1990 &
 1990 & 1994 \\ \hline
Expt. &\hspace{0.1in} -2.22 & -8.48 & -28.3 & -27.2 & -25.8 & -32.0 \\
Theory &\hspace{0.1in} -2.22 & -8.47(2) & -28.3(1) & -26.5(2) & -25.7(2)
& -32.4(9) \\ \hline
\end{tabular}

\end{table}

We can also extract from these results the (average) amounts of potential
energy accruing from 2Nf and 3Nf, and an upper limit estimate of 4Nf from the
error bar on the $\alpha$-particle energy:
$$
\langle V_{NN} \rangle \sim 20\, {\rm MeV/pair} \, , \eqno (2a)
$$
$$
\langle V_{3Nf} \rangle \sim 1\, {\rm MeV/triplet} \, , \eqno (2b)
$$
$$
\langle V_{4Nf} \rangle \lsim \, 0.1\, {\rm MeV/quartet} \, . \eqno (2c)
$$
This geometric progression is in accordance with power-counting predictions.
We emphasize again that weak many-nucleon forces are essential for tractability.
Although we can use the vast amount of $NN$ scattering data to fit the 2Nf, we
have no such options for the 3Nf.

As we stated earlier, the long-range two-pion-exchange 2Nf is weak because of
CS. There have been a number of recent papers\cite{9} treating aspects of the
problem. Ord\'o\~nez et al. was the first to develop a chiral force. Ballot et
al. shows how CS arranges cancellations to keep the force weak. Because this
potential is relatively weak, no experimental demonstration of its existence yet
exists. 

\section{Effective Field Theories and Isospin Violation in the NN Force}

We have already mentioned effective field theories several times. These
theories\cite{10} can be viewed as approximations to a known theory (such as
QCD) or to an unknown (as yet) theory that is valid to a larger energy scale.
Such theories are typically non-renormalizable and can even be nonrelativistic
(or semirelativistic). A price is paid for the lack of renormalizability. As the
order of the calculation increases, more and more parameters appear that must be
fit to data, making higher-order calculations both more difficult and less
predictive. What saves the scheme is that for sufficiently low energies the
expansion is a series in $\bar{p}/\Lambda$, and should converge fairly rapidly
in most circumstances. 

Why is this better than simply using a model? Models typically have fewer 
parameters, for example. The strengths of this scheme are at least threefold:

\begin{itemize} 
\item The individual terms in the Lagrangian are based on symmetry (e.g., CS). 
\item It is {\bf not} a model -- at low energies it should have all of the 
content of the original or covering theory (but less predictive power).
\item Power counting allows one to estimate with reasonable accuracy the size 
of various contributions without detailed calculations. 
\end{itemize} 
The latter is an extremely useful and powerful technique, as we have seen.

As an illustration we will review the work of van Kolck\cite{11} on isospin
violation in the nuclear force, and estimate the size of isospin violation in
one-pion-range nuclear forces by comparing to results from the Nijmegen PSA. In
the latter work, after accounting for well-established forms of isospin
violation such as the $NN$ Coulomb force, magnetic moment interactions between
the nucleons, the $n-p$ mass difference in the nucleon's kinetic energy, etc.,
three $\pi N$ coupling constants were determined\cite{4}. These correspond to
$\pi^0$-exchange between two protons, $f^2_{\pi^0 pp}$, or the exchange between
a neutron and proton of a $\pi^0$, $f_{\pi^0 pp}f_{\pi^0 nn}$, or a charged
pion, $f^2_{\pi^c np}$. OPEP depends linearly on $f^2$, which we define as 
$$
f^2 = \frac{1}{4 \pi} \left(\frac{g_A \, m_{\pi^+} \, d}{2 f_{\pi}} 
\right)^2 \, , \eqno (3)
$$
where $g_A$ is the axial-vector coupling constant and $d-1$ is the
Goldberger-Treiman(GT) discrepancy ($>$ 0) and a measure of chiral-symmetry
breaking. In terms of $d$ and $G \sim 13$ (the ``pseudoscalar'' form of the $\pi
N$ coupling constant) we write the GT relation\cite{12} in the form 
$$ 
\frac{G}{M} = \frac{g_A \, d}{f_{\pi}} \, . \eqno (4)
$$
For reference purposes we note that setting d to 1 and using current values of 
$g_A$ and $f_{\pi}$ produces $f^2 (d=1)$ = 0.0718(5) in eq. (3).

\begin{table}[hbt]
\centering
\caption{Pion-nucleon coupling constants determined by the 
Nijmegen\protect\cite{4} PSA.}

\begin{tabular}{|ccc|c|}
\hline
$f^2_{\pi^0 pp}$ & $f_{\pi^0 pp} f_{\pi^0 nn}$ & $f^2_{\pi^c np}$ & Type \\ 
\hline \hline
0.0751(6) & 0.0752(8) & 0.0741(5) & $np$ and $pp$ \\
         & 0.0745(9) & 0.0748(3) & $np$ only \\ \hline
\end{tabular}
\end{table}

Two sets of coupling constants are available and are shown in Table 4.1. The
combined $np$ and $pp$ data sets yield the three values in the first line, while
a preliminary solution for $np$ scattering alone gives the two values in the
second line. These values are all roughly 4\% greater than the value of 
$f^2 (d=1)$, implying that $d-1$ is $\sim $ 2\%, much lower than most previous 
values. We note that systematic effects would enter these results in different 
ways. The $pp$ data must be carefully corrected for the Coulomb interaction, 
which plays only a very minor role in $np$ scattering. We also note that 
charge-symmetry breaking (CSB) in the pion-range force involves a $pp-nn$ \
comparison and does not contribute in the second ($np$ only) case, but charge 
dependence (CD) will. In qualitative terms our three experimental numbers from 
Nijmegen determine the isospin-symmetric $d$ and the CSB and CD $\pi N$ 
coupling constants.

The formalism for analyzing these results has been developed by van
Kolck\cite{11}. Isospin violation (IV) can be divided into 3 convenient types
based on its origin. The u-d quark-mass difference generates one type of IV that
has the (tensorial) character of the third component of total isospin, and
should be proportional to $\epsilon m_{\pi}^2$, where $\epsilon = \frac{m_d -
m_u}{m_d + m_u} \sim 0.3$ characterizes the quark masses. These contributions
are also chiral-symmetry breaking. Hard electromagnetic (EM) processes at the
quark level will have a different (tensorial) isospin character and generate the
second type, which should be proportional to $\bar{\delta} m_{\pi}^2$, the
(squared) difference of the pion masses. Soft EM processes form the third
category and are of order $\alpha$, the fine structure constant. There are a
daunting number of individual diagrams that must be calculated\cite{13} and we
will present here only a much simplified overview of where individual mechanisms
(most of which are well known) fit into this scheme. 

The first difficulty is that we must try to place these categories into an
approximate (size) relationship with each other. Van Kolck has argued that since
the pion-mass difference is primarily of type II (i.e., EM) and the nucleon-mass
difference mostly type I (i.e., quark-mass difference), these terms should be
considered on the same level. Although this prescription probably underestimates
most EM contributions slightly, it results in the entries in Table 4.2. The
order refers to power counting, with OPEP roughly of order 0, while tree and
loop refer to the structure of the corresponding diagrams, such as those in Fig.
(3). The upper entry in each box describes the fundamental interaction, while
the lower (bold-faced) entry describes its effect on the nuclear dynamics. 

\begin{table}[hbt]
\centering
\caption{Contributions to isospin violation in the NN force in orders
1,2, and 3 beyond the usual OPEP, including tree and loop processes, are
shown in individual boxes. The first entry of each type corresponds to the 
elementary vertex, while directly below the single line (in boldface) is 
the corresponding nuclear contribution. Charge-dependent and 
charge-symmetry-breaking processes are denoted CD and CSB, respectively.}
\begin{tabular}{|c||l|l|}
\hline
{ORDER} & {QUARK MASSES} & {ELECTROMAGNETIC}  \\ \hline \hline
{}&{}&Coulomb interaction \\
1-tree&Nucleon masses&Pion masses \\ \cline{2-3}
{}&{}&{\bf pp Coulomb force} \\ 
{}&{\bf Overall energy shift}&{\bf OPEP(masses) [CD]}\\ \hline \hline
{}&$\pi$N interaction&Nucleon masses\\
2-tree&Pion masses&{}\\ \cline{2-3}
{}&{\bf OPEP($\mbox{\boldmath $\pi$}_0$) [CSB]}&{\bf Overall energy shift}\\
{}&{\bf OPEP(masses) [CD]}&{}\\ \hline \hline
{}&Nucleon kinetic energies&Other nucleon EM\\
3-tree&{}&$\pi$N interaction\\ \cline{2-3}
{}&{\bf Nucleus kinetic energy}&{\bf Breit interaction}\\
{}&{}&{\bf OPEP($\mbox{\boldmath $\pi$}_0$) [CD + CSB]}\\ \hline \hline
{}&Nucleon masses and KE&Nucleon masses and KE\\
{}&Pion masses - OPEP&Pion masses  - OPEP\\
3-loop&OPEP($\pi_0$) [CSB]&{\bf OPEP($\mbox{\boldmath $\pi$}_0$) [CD} and CSB]\\
{}&{}& $\mbox{\boldmath $\pi$}-\mbox{\boldmath $\gamma$}$ {\bf NN forces [CD]}\\
\hline \hline
\end{tabular}

\end{table}

Some examples can be correlated with entries in Fig. (3). Isospin violations are
indicated by a cross, so that Fig. (3a) indicates the CD effect of the pion-mass
difference on OPEP. This is listed in the first box (1-tree) together with the
$pp$ Coulomb force shown in Fig. (3b). Smaller EM effects (magnetic moment
interactions, etc.) are listed in the third box under Breit interaction. The
shift of nucleon masses ($\delta M_N \sim M_n -M_p$) has three consequences: (1)
an overall energy shift that does not affect physics; (2) a change in the
kinetic energies, $p^2/2M$, of the individual nucleons according to whether they
are neutrons or protons (indicated in the third box); (3) the effect of the mass
difference inside loops. The nucleon-mass splitting is proportional to $t_z$,
the third component of total (nucleon) isospin, which equals (Z-N)/2 and is
fixed. On the other hand in loops where pions are circulating it is possible for
an $np$ pair to virtually dissociate into $pp\pi^-$. This changes the {\bf
nucleons'} isospin and demonstrates that loops with different nucleon masses can
in principle play a role in IV. 

\begin{figure}[htb]
 \epsfig{file=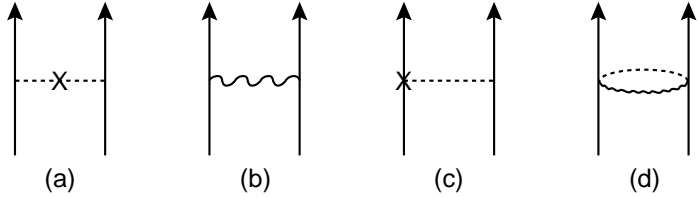,height=1.0in,bbllx=90pt,bblly=492pt,bburx=461pt,bbury=582pt}
 \caption{Nuclear isospin-violating interactions, with the pion-mass-difference
effect on OPEP shown in (a), the static Coulomb interaction in (b), the 
isospin-violating pi-nucleon coupling illustrated in (c), and the double-seagull
(transverse-)photon-pion exchange sketched in (d).  Solid lines are
nucleons, dashed lines are pions, small wavy lines are (virtual) transverse
photons, while large wavy lines are Coulomb interactions.}
\end{figure}

All of the effects listed above (except loop contributions) are included in the
Nijmegen analysis, and we don't need to correct for them. Two additional
categories do play a role: the effect of isospin violation in the $\pi^0 N$
coupling constants (indicated by the cross in Fig. (3c)), and $\pi-\gamma$ range
$NN$ forces\cite{14} of the type shown in Fig. (3d) and the fourth box in Table
4.2. The latter are not included in analyzing $NN$ scattering data and their
effect is unknown, although their order is the same as CD modifications of
$f^2$. 

Ignoring loops and terms that contribute to them, we can write van Kolck's
chiral-symmetry-breaking and isospin-violating terms that contribute at 
tree level in the form:
$$
L \sim - \frac{1}{f_{\pi}}\, \overline{N} \mbox{\boldmath$\sigma$} \cdot 
\mbox{\boldmath$\nabla$} [g_A\, d\, \mbox{\boldmath$t$} \cdot 
\mbox{\boldmath$\pi$} -\frac{\beta_1}{2} \pi_0 -\frac{\bar{\beta}_{10}}{2} 
t_z \pi_0 ] N \, . \eqno (5)
$$
The three terms in order are the usual $\pi N$ coupling (including the GT 
discrepancy) and the CSB and CD $\pi N$ vertices. This form can be analyzed with
the Nijmegen coupling constants to produce the results in Table 4.3. We had
previously estimated $d-1$; the values of $\beta_1$ and $\bar{\beta}_{10}$ are
consistent with zero and there is no evidence of any systematic disagreement
between the two sets of results. We note that the dimensional-power-counting
estimate of $\beta_1$ is $\sim 6 \cdot 10^{-3}$, while that of
$\bar{\beta}_{10}$ is $\sim 2 \cdot 10^{-3}$. Thus, our results (consistent with
zero) are also consistent with expectations of a very small isospin violation.

\begin{table}[hbt]
\centering
\caption{GT discrepancy and isospin-violating pion-nucleon coupling constants 
determined by the Nijmegen PSA.}

\begin{tabular}{|ccc|c|}
\hline
$d-1$ & $\beta_1$ & $\bar{\beta}_{10}$ & Type \\ 
\hline \hline
1.6(5)\% & 1(9) $\cdot 10^{-3}$ & -19(16)$ \cdot 10^{-3}$ & $np$ and $pp$ \\
2.1(5)\% &                      &   5(16)$ \cdot 10^{-3}$ & $np$ only \\ \hline
\end{tabular}

\end{table}

\section{Summary}
Chiral symmetry arises at the quark level in QCD and persists in descriptions
based on pions and nucleons as effective degrees of freedom. One-pion exchange
dominates in the binding of light nuclei and in meson-exchange currents. Chiral
symmetry provides order in nuclear forces:  without this symmetry nuclear
physics would be intractable. Turning the argument around, the tractability of
nuclear physics provides strong evidence for chiral symmetry, which weakens
$N$-body forces as $N$ increases and $n$-pion exchanges compared to OPEP.
Isospin-violation upper limits in OPEP obtained from the Nijmegen PSA are
compatible with dimensional-power-counting estimates. Finally, few-nucleon
systems continue to be the testing ground for new ideas in nuclear physics
because of our ability to calculate accurately in those systems. 

\section{Acknowledgments}
This work was performed under the auspices of the U.\ S.\ Department of Energy. 
J.\ de Swart and R.\ Klomp generously provided preliminary results for 
our analysis.

\SaveFinalPage
\end{document}